\documentclass[a4paper, 12pt]{article}

\usepackage{graphicx}
\usepackage{url}
\usepackage{amsbsy,amssymb} 
\usepackage{amsmath}
\usepackage{abstract}
\usepackage{hyperref}
\usepackage{color}

\usepackage{amsmath,amsthm}

\newtheorem{Corollary}{Corollary}

\newtheorem{Theorem}{Theorem}
\newtheorem*{Proof}{Proof}
\newtheorem{Lemma}{Lemma}
\newtheorem{Definition}{Definition}
\newtheorem{Construction}{Construction}

\begin{document}

{\LARGE\centering{\bf{Landauer's principle as a special case of Galois connection}}}

\begin{center}
\sf{Rados\l aw A. Kycia$^{1,2,a}$}
\end{center}

\medskip
\small{
\centerline{$^{1}$Masaryk Univeristy}
\centerline{Department of Mathematics and Statistics}
\centerline{Kotl\'{a}\v{r}sk\'{a} 267/2, 611 37 Brno, The Czech Republic}
\centerline{\\}
\centerline{$^{2}$Cracow University of Technology}
\centerline{Faculty of Physics, Mathematics and Computer Science}
\centerline{Warszawska 24, Krak\'ow, 31-155, Poland}
\centerline{\\}

\centerline{$^{a}${\tt
kycia.radoslaw@gmail.com}}
}

\begin{abstract}
\noindent
It is demonstrated how to construct a Galois connection between two related systems with entropy. The construction, called the Landauer's connection, describes coupling between two systems with entropy. It is straightforward and transfers changes in one system to the other one preserving ordering structure induced by entropy. The Landauer's connection simplifies the description of the classical Landauer's principle for computational systems. Categorification and generalization of the Landauer's principle opens area of modelling of various systems in presence of entropy in abstract terms.
\end{abstract}

\section{Introduction}

There are various kinds of entropy describing different systems, e.g. in  computations, physics, dynamical systems. In continuous thermodynamic system, e.g., ideal gases, the entropy has precise meaning of a function which provides foliation of the thermodynamic space of states \cite{AxiomaticThermodynamics, GeometryOfPhysicsFrankel}, which is the statement of the Caratheodory formulation of the second law of thermodynamics. This approach requires a continuous structure on the space of states of the system \cite{EntropyOrdering, EntropyOrdering2}. Entropy can be used as comparison measure between states \cite{EntropyOrdering, EntropyOrdering2}, which will be useful later in the paper. There is also a point of view that the entropy in a theory can be traced to inaccurate (as it always is) measurement \cite{EntropyLychagin}, and the only crucial thing is the difference in entropy and not the entropy itself. In the theory of dynamical systems topological entropy is used to measure the level of dynamical complexity of a system \cite{Katok}. In information theory the (Shanon) entropy measures how information is produced by its (stochastic) source \cite{InformationEntropy}. This discussion can be largely extended, however it is not the aim of the paper to make extensive research on the vast literature of the subject. These  various interpretations show that the notion of entropy is not well understood.

Apart from these different approaches, better insight is possible when systems with entropy are 'connected' in the following sense. In 1961 Rolf Landauer introduced the principle (Landauer's principle) in irreversible computing in which he postulated that every act of erasing information results in expelling at least $Tk_{B}\ln(2)$ [Joule per bit] (here $k_{B}$ is the Boltzmann constant, $T$ is the temperature of the system) heat to the environment \cite{Landauer}, i.e.  increases thermodynamic entropy. This principle has profound implication in explaining old thermodynamic paradox of the Maxwell's demon \cite{Bennet, BennetDemon}. There was a dispute on the validity of this principle, however careful derivation \cite{LandauerExplained} and experimental results, e.g. experimental setup close to the one proposed by Landauer that is presented in \cite{LandauerMeasurment}, and recent verification in quantum systems that can be found in \cite{QuantumLandauer}, prove that the principle is correct.
The principle is based on the assumption that every computational system (in principle computer memory) is implemented with the help of physical system and this is a link between information and physical realms.

This paper is an attempt to generalize this principle to every system that contains entropy. This generalized connection between systems is the minimal that preserves entropy-induced ordering. It will be described what kind of structures from category theory \cite{CategoryTheoryForWorkingMathematicans, CategoryGentleIntroduction} are involved in the Landauer's principle. Category theory approach was used in studying entropy, e.g. in \cite{BaezEntropy1, BaezEntropy2}. In this view it is an extension of the paper \cite{LandauerExplained}, where this categorical viewpoint was abandoned. It will be shown that Tab. 1 of \cite{LandauerExplained} (see Tab. \ref{Tab1_FormPaperLandauer} in this paper) is an indication of the Galois' connection.
\begin{table}[!h]
\centering
 \begin{tabular}{|c|c|c|}
 \hline
 Possibilities &  \begin{tabular}{@{}c@{}}Thermodynamically \\ reversible\end{tabular}  & \begin{tabular}{@{}c@{}}Thermodynamically \\ irreversible\end{tabular} \\ \hline
 Logically reversible & YES  & YES \\ \hline
 Logically irreversible & NO & YES \\ \hline
\end{tabular}
\caption{Table 1 from \cite{LandauerExplained} defining connection between memory operations and their realizations in thermodynamic system. For definition of (ir)reversibility see below.}
\label{Tab1_FormPaperLandauer}
\end{table}
The mappings (functors) between states of one system (e.g. a logical system) and the second one (a thermodynamic realization of the logical system) preserve entropy properties. These features, when properly defined, are exactly properties required for the existence of a Galois connection between these two systems seen as ordered sets.

A few remarks are in order before we provide details. The category theory is not an alternative approach in proving the Landauer's principle, in the same way as it is not a tool to prove basic properties of  objects in mathematics. Instead, it offers a layer of abstraction (called 'abstract nonsense') that allows to promote some specific features (e.g. the Landauer's principle proven by thermodynamic methods) to  'universal properties' that can be observed in any other system that shares specific common features. In order to use this link between specific phenomena/object and category theory these universal properties have to be proven using specific domain methods. Then the language of category theory can be used to prove and understand even more on abstract level. The link or 'a bridge' between original Landauer's principle and the Galois connection for systems with entropy will be formulated and proven in Theorem \ref{Theorem_Main} below.

This approach is from bottom to top and recently there is some trend in applied science to use such kind of abstract approach to concrete models \cite{SpivakSketches}, especially adjoin functors\footnote{``Adjoint functors arise everywhere.'' S. Mac Lane} in physics \cite{AdjointInPhysics} of which the Galois connection is a special and distinguished case.

This paper is organized as follows: In the next section a brief overview of the mathematical notion of entropy in thermodynamics and the Galois connection is presented for the reader's convenience. Following section contains the definition of the Landauer's connection that relates systems with entropy. Then some various examples of the Landauer's connection, including description of the classical Landauer's principle in terms of the Landauer's connection will be outlined. The paper concludes with the discussion on possible implications. First part of the paper is strict and precise, however the Examples section varies in the level of precision since the description of complicated system on high level of generality is in principle impossible or depends on too many details to include them here. Therefore many conclusions in that section should not be taken too strict, and are in fact hypotheses or general features rather than strict claims. We however believe that it is worth to include them here as they illustrate wide range of disciplines in which the generalized Landauer's principle can be possibly applied.

\section{Related work}
This section summarize important facts from entropy theory of thermodynamic systems (based on \cite{EntropyOrdering, EntropyOrdering2}), and the definition and properties of the Galois connection (mainly following \cite{CategoryGentleIntroduction}).

\subsection{Entropy and ordering in thermodynamics}
\label{Appendix_Entropy}

This subsection presents the poset (pre-ordered set) structure constructed on the state-space of thermodynamic system, however some parts are also valid for other types of systems with entropy. We will closely follow \cite{EntropyOrdering, EntropyOrdering2}.

A state of thermodynamic system is associated with a point (equilibrium state) $X$ in the state-space $\Gamma$. This space can be topologized and coordinates, that describe physical quantities like energy, volume etc., can be introduced, however we do not need it for what follows (for details see \cite{EntropyOrdering2}). Equilibrium state is attained when system is left to itself. 

Crucial operation that can be introduced on the thermodynamic system is the scaling $\Gamma^{\lambda}$, or for the state - $\lambda X$, which is an action of abelian multiplicative group $\lambda \in (\mathbb{R}_{+}\cup \{0\},\cdot)$ on the state-space. It fulfils composition law $(\Gamma^{\lambda})^{\mu}=\Gamma^{\lambda \mu}$ and $\mu (\lambda X)=(\mu\lambda)X$, with obvious unity $\Gamma^{1}=\Gamma$ and $1X=X$. This scaling multiplies extensive properties (coordinates) of the system and do not alter intensive ones. It allows us to build bigger system from the small similar pieces. Similar construction is a composition of two systems $X \in \Gamma_{1}$ and $Y \in \Gamma_{2}$, that is $(X,Y)\in \Gamma_{1} \times \Gamma_{2}$.

{\it Adiabatic transition/process} is a change of state which is done without influence from outside of the system. It can occur abruptly or slowly. This allows us, according to \cite{EntropyOrdering}, to introduce partial ordering as follows: If $Y$ can be reached by an adiabatic process (is {\it adiabatically accessible}) from the state $X$, then we denote it as
\begin{equation}
 X \preccurlyeq Y.
\end{equation}
We can define adiabatic equivalence (which is antisymmetric relation $\preccurlyeq$) saying that $X$ is {\it adiabatically equivalent} with $Y$, in symbols $X \sim Y$, if $X \preccurlyeq Y $ and $Y \preccurlyeq X$. The classes of equivalence are called {\it adiabats}. Obviously, there is reflexivity of $\preccurlyeq$ since identity process is also an adiabatic process - when the system is in equilibrium.

If there is no symmetry between $X$ and $Y$ we say that $X \prec\prec Y$ if $X \preccurlyeq Y$  and $Y \nprec X$. In this case the (isolated) transition between these states is called {\it irreversible adiabatic process}.

Two states are {\it comparable} if there is $\preccurlyeq$ relation between them. Not every pair of states is comparable since, e.g., they have different chemical composition. In addition, the relation $\preccurlyeq$ is transitive, i.e., if $X \preccurlyeq Y$ and $Y\preccurlyeq Z$ then $X\preccurlyeq Z$. Reflexivity, antisymmetry and transitivity shows that the relation $\preccurlyeq$ is a partial order on $\Gamma$.

We want to transfer the ordering to the ordering of the real line. This can be done using entropy function:
\begin{Definition}\cite{EntropyOrdering, EntropyOrdering2}
\label{Def.Entropy}
This order structure induces the entropy function which is a mapping $S: \Gamma \rightarrow \mathbb{R}$ that fulfils:
\begin{itemize}
 \item {{\it Monotonicity}: For two comparable states $X$ and $Y$
  \begin{equation}
    X \preccurlyeq Y \quad \Leftrightarrow \quad  S(X) \leq S(Y);
  \end{equation}
 }
 \item{{\it Additivity}: For two states $X$ and $Y$ the entropy of compound state $(X,Y)$ is
  \begin{equation}
   S(X,Y) = S(X) + S(Y);
  \end{equation}
 }
 \item{{\it Extensivity}: For $\lambda >0$ and a state $X$
   \begin{equation}
    S(\lambda X) = \lambda S(X).
   \end{equation}
 }
\end{itemize}
\end{Definition}

The properties of entropy allows us to transfer the relation $\preccurlyeq$ to the $\leq$ relation on the real line. Therefore, the above definitions of processes can be rewritten as:
\begin{itemize}
 \item {{\it Reversible adiabatic process}: 
 \begin{equation}
  X \sim Y \quad \Leftrightarrow \quad S(X) = S(Y);
 \end{equation}
  }
 \item {{\it Irreversible adiabatic process}: 
 \begin{equation}
  X \prec \prec Y \quad \Leftrightarrow \quad S(X) < S(Y).
 \end{equation}
 }
\end{itemize}

If, in addition, all states are comparable in state-space $\Gamma$ (Comparison Hypothesis \cite{EntropyOrdering, EntropyOrdering2}), then there exists unique, up to affine transformation, entropy, that fulfils above properties for $\preccurlyeq$ and supplied with some additional conditions (see axioms A1-A6 of \cite{EntropyOrdering}). This property says that $\preccurlyeq$ is total order on $\Gamma$ and $S$ function transfers total order on $\Gamma$ to the total order on $\mathbb{R}$. We will be assuming Comparison Hypothesis.

These ingredients are minimal for our purposes. The reader interested in full axiomatization of entropy is refereed to \cite{EntropyOrdering, EntropyOrdering2} and references therein.

Most of the properties presented here holds for different types of entropy, not necessary thermodynamic one.

\subsection{Galois connection}
\label{Appendix_GaloisConnetion}

Let us also recall the definition of the Galois connection between two posets following \cite{CategoryGentleIntroduction}, and a few its properties from the vast literature on this subject, e.g., \cite{GaloisConnectionDefinition1, GaloisConnectionDefinition2, GaloisConnectionDefinition3, GaloisConnectionDefinition4, CategoryTheoryForWorkingMathematicans, CategoryGentleIntroduction}.

First, the three definitions of functors that preserve ordering have to be given \cite{CategoryGentleIntroduction}. Let $\mathcal{C}=(C, \preccurlyeq)$ and $\mathcal{D} = (D,\sqsubseteq)$ are two posets\footnote{Here $\preccurlyeq$ and $\sqsubseteq$ are partial order relations.} then the mapping (functor) $F:\mathcal{C}\rightarrow \mathcal{D}$ is
\begin{itemize}
 \item {a {\it monotone} if for any $x,y \in C$, if $x \preccurlyeq y$, then $Fx \sqsubseteq  Fy$;}
 \item {an {\it order-embedding} if for all $x,y \in C$, $x \preccurlyeq y \Leftrightarrow Fx \sqsubseteq Fy$;}
 \item {an {\it order-isomorphism} iff $F$ is surjective order-embedding;}
\end{itemize}

Next, following \cite{CategoryGentleIntroduction}, Definition 116, the Galois connection is given as
\begin{Definition} 
\label{Def.GaloisConnection}
\cite{CategoryGentleIntroduction}
 Suppose that $\mathcal{C}=(C, \preccurlyeq)$ and $\mathcal{D} = (D,\sqsubseteq)$ are two posets, and let $F:\mathcal{C}\rightarrow \mathcal{D}$ and $G:\mathcal{D}\rightarrow \mathcal{C}$ be a pair of functors such that for all $c \in C$, $d\in D$,
 \begin{equation}
  Fc \sqsubseteq d \quad \Leftrightarrow \quad c \preccurlyeq Gd.
  \label{Eq.GaloisConnectionG}
 \end{equation}
Then $F$ and $G$ form a {\it Galois connection} between $\mathcal{C}$ and $\mathcal{D}$. When this holds, we write $F\dashv G$, and $F$ is said to be the {\it left adjoint} of $G$, and $G$ is the {\it right adjoint} of $F$.
\end{Definition}

There are alternative conditions to (\ref{Eq.GaloisConnectionG}), which are given by (Theorem 144 of \cite{CategoryGentleIntroduction})
\begin{Theorem}\cite{CategoryGentleIntroduction}
\label{Th.GaloisConnectionEquivalentConditions}
 In the assumptions of Def. \ref{Def.GaloisConnection}, $F \dashv G$ if and only if
 \begin{enumerate}
  \item {$F$ and $G$ are both monotone, and}
  \item {for all $c\in C$, $d\in D$, $c\preccurlyeq GFc$ and $FGd \sqsubseteq d$, and}
  \item {$FGF=F$ and $GFG=G$.}
 \end{enumerate}
\end{Theorem}

\section{Main results}
In this section we use the properties of entropy \cite{EntropyOrdering, EntropyOrdering2} and the Galois connection \cite{CategoryGentleIntroduction} to construct connection between systems with entropy. We will consider general entropy, and not necessary thermodynamic one.

The plan of deriving the Galois connection from entropy consists of the following steps:
\begin{enumerate}
 \item { state-space (G-Set) $+$ entropy $\rightarrow$ total ordering,}
 \item { total ordering $\rightarrow$ poset (G-poset) structure,}
 \item { two posets $\rightarrow$ Galios (Landauer's) connection between them.}
\end{enumerate}

\textbf{Step 1.} The main point is to introduce state-space set $\Gamma$. In the case of thermodynamic systems the scaling of the system is modelled as an action of the multiplicative group $(\mathbb{R}^{+},\cdot,1)$ on the set $\Gamma$, which preserves ordering (as described in \cite{EntropyOrdering, EntropyOrdering2} and Section \ref{Appendix_Entropy}). Such kind of element $\{\Gamma,(\mathbb{R}^{+},\cdot,1)\}$  is the object of a G-Set category \cite{GSet}, namely $\mathbb{R}^{+}$-Set category. However in non-thermodynamic systems (e.g. information theory) there is usually no group action and we have the following options: either narrow description of state-space to the category Set, select the category of G-Set with the trivial group $(1,\cdot,1)$, or modelled state-space on the $\mathbb{R}^{+}$-Set category with trivial group action. We select the last possibility since it gives a more uniform approach.
\begin{Definition}
System space is the object of G-Set category, i.e. $\{\Gamma,(\mathbb{R}^{+},\cdot,1)\}$, where the multiplicative group acts on state-space $\Gamma$.
\end{Definition}

The second ingredient is an entropy function $S:\Gamma \rightarrow \mathbb{R}$. For example for thermodynamic systems the entropy must fulfil properties of Definition \ref{Def.Entropy}.

It is assumed that every point of $\Gamma$ is in the domain of the entropy function. In thermodynamic systems this assumption is called Comparison Hypothesis and not always is true as described in \cite{EntropyOrdering, EntropyOrdering2}. However we will assume it holds (e.g. no systems with chemical reactions for thermodynamic systems). 

The existence of entropy allows us to define 
\begin{Definition}
Total ordering $\preccurlyeq$ on $\Gamma$ is defined in the following way
\begin{equation}
    X \preccurlyeq Y \quad \Leftrightarrow \quad  S(X) \leq S(Y),
\end{equation}
for $X,Y \in \Gamma$. 
Likewise, 
\begin{equation}
 X=Y \quad \Leftrightarrow \quad  S(X) = S(Y).
\end{equation}
\end{Definition}

The above construction from entropy to ordering for thermodynamic systems is the reverse of the argument from \cite{EntropyOrdering, EntropyOrdering2}, and it is also sketched in Section \ref{Appendix_Entropy}.

\textbf{Step 2.} 
The existence of total ordering $\preccurlyeq$ on $\Gamma$ allows us to define a poset structure. However for accounts of additional group structure of scaling, the more general approach will be to use  G-Pos category \cite{GPoset}, i.e. posets with a group action. The group action is needed only when the scaling is present in the system (in particular thermodynamics). In other systems without scaling the group action is trivial. Therefore we will omit the group action/scaling part when it is not important in the context and provide modifications in the presence of G-Pos structure of group action/scaling later.
\begin{Definition}
 The {\it entropy system} is the object of G-Pos category, which objects are $\mathcal{G}=(\Gamma, \preccurlyeq)$, with preserving ordering group $(\mathbb{R}^{+},\cdot,1)$ action\footnote{If for $X,Y \in \Gamma$ there is $X\preccurlyeq Y$, then for $\lambda \in \mathbb{R}^{+}$ there is $\lambda X \preccurlyeq \lambda Y$.}, where the (partial or) total order is given by the entropy function $S:\Gamma \rightarrow \mathbb{R}$.
\end{Definition}

\textbf{Step 3.} The third key ingredient to formulate the Landauer's connection is the Galois connection from category theory. The short overview of its definition and basic properties are collected in Section \ref{Appendix_GaloisConnetion} for the reader's convenience.

The definition of the Galois connection suggest that it can relate two thermodynamic or, more generally, entropy systems with state-spaces $\Gamma_{1}$ and $\Gamma_{2}$. In such case the existence of entropy imposes poset structure (we omit group action structure for clarity for the moment). This gives our main observation - the following definition is reformulation of the Galois connection (Definition \ref{Def.GaloisConnection}) in terms of the entropy, and from the historical reasons we call it  the Landauer's connection (or generalized Landauer's principle). In the below definition every poset is treated as a category on its own.
\begin{Definition} \label{Def.LandauerConnection} \textbf{The Landauer's connection and Landauer's functor} \\
 Entropy system $\mathcal{G}_{1}=(\Gamma_{1}, S_{1} )$ is implemented/realized/simulated in the entropy system $\mathcal{G}_{2}= (\Gamma_{2}, S_{2})$ when there is a Galois connection between them, namely, there is a functor $F:\mathcal{G}_{1} \rightarrow \mathcal{G}_{2}$ and a functor $G:\mathcal{G}_{2} \rightarrow \mathcal{G}_{1}$ such that $F \dashv G$. 
 
 In terms of the entropy the condition (\ref{Eq.GaloisConnectionG}) is given as
 \begin{equation}
  S_{2}(Fc) \leq S_{2}(d) \Leftrightarrow S_{1}(c) \leq S_{1}(Gd).
  \label{Def.LandauersConnection}
 \end{equation}
We name the functors $F$ and $G$ the Landauer's functors.

In case when the group action on set-state is nontrivial, i.e. when scaling of states is present, then we operate on G-Posets and in such situation every functor above, say $\tilde{F} = (F,\phi)$ consists of two parts 
\begin{itemize}
 \item {Set part of the functor: $F:\Gamma_{1} \rightarrow \Gamma_{2}$,}
 \item {A function $\phi$ that is a surjective group endomorphism of $(\mathbb{R}^{+},\cdot,1)$,}
\end{itemize}
that acts as $\tilde{F}(\lambda X) = \phi(\lambda)FX$.
\end{Definition}
We intuitively explain that the functorial properties holds. In the case when there is no scaling ($\phi$ is trivial), if there are mappings $f,g:\Gamma_1 \rightarrow \Gamma_1$ then they induce the mappings $Ff, Fg: F\Gamma_2 \rightarrow F\Gamma_2$ on the connected system and the composition $g \circ f$ is mapped into $F(g\circ f) = Fg \circ Ff$. In addition, if there is no transition changing the entropy in $\Gamma_{1}$, it corresponds to the identity mapping on $\Gamma_{1}$ and this corresponds to the identity mapping on $F\Gamma_{1}$. Similarly for $G$ functor.

Definition \ref{Def.LandauerConnection} is reasonable as from the first condition of Theorem \ref{Th.GaloisConnectionEquivalentConditions} from Section \ref{Appendix_GaloisConnetion}, the realization of $\mathcal{G}_{1}$ on $\mathcal{G}_{2}$ preserves ordering. From the second condition we get that the mapping $GF$ and $FG$ does not give lower and respectively higher entropy states than the original states, and the third condition shows that $FG$ and $GF$ preserves image of $F$ and $G$ respectively. For thermodynamic systems the Landauer's functors preserve entropy properties given in Definition \ref{Def.Entropy}. Therefore for such simulation the Landauer's connection is needed, as it is minimal order preserving connection between two posets, and therefore entropy systems.

Surjectivity of $\phi$ is only a technical assumption that simplifies what follows. This assumption is added only for removing additional degree of freedom, since we can always choose group action to compensate.

It is obvious that when one system has scaling and the other has not (trivial action of the group) then the mapping $\phi$ is the trivial map. However for nontrivial group action we have the following
\begin{Corollary}
For Landauer's connected functors in the presence of scaling, if $\tilde{F} \dashv \tilde{G}$, then $\tilde{F} = (F, \phi)$, $\tilde{G}=(G,\phi^{-1})$ and $F \dashv G$. Therefore $\phi$ is an isomorphism of groups.
\end{Corollary}
\begin{Proof}
 Assume that $\tilde{G}=(G,\psi)$ for the moment. In the relations $S_{2}(FGFc)=S_{2}(Fc)$, and $S_{1}(GFGd)=S_{2}(Gd)$ substitute $c \rightarrow \lambda c$ and $d \rightarrow \gamma d$, where $\lambda, \gamma \in \mathbb{R}^{+}$. Then one gets  $\psi \circ \phi \circ \psi (\lambda) = \psi(\lambda)$ and likewise $\phi \circ \psi \circ \phi (\gamma) = \phi(\gamma)$. Since $\psi$ and $\phi$ are surjective (group homomorphisms), therefore $\phi^{-1} = \psi$.
\end{Proof}

The word 'realization' or 'simulation' explains that usually we are interested in simulating one, possibly abstract system, which consists some kind of entropy that introduces poset structure in its state-space, e.g. binary computations, using its physical implementation in terms of electronic system, spin system, or any other computing realization, where thermodynamic entropy is given. We can also consider simulation of physical system by the other physical system. If the Landauer's connection is present the simulated system will behave as its connected counterpart, following the Second Law of Thermodynamics.  If the physical system as one of the Galois connected category is involved, then the connection transfers the Second Law of Themrodynamics to the other category which does not have to be connected with physical world (e.g. from electronic circuits to computation described by the Shanon entropy). This issue will be described in details in the next subsection.

Reformulation of the properties of Landauer's functors is given by
\begin{Corollary}
\label{Corollary_LandauerFunctorsEquivalentDef}
 The equivalent conditions to (\ref{Def.LandauersConnection}) are as follows
\begin{enumerate}
 \item {For $c_{1},c_{2}\in C$, if $c_{1} \preccurlyeq c_{2}$ then $S_{2}(Fc_{1}) \leq S_{2}(Fc_{2})$;  analogously for $G$ functor. In other words, $F$ and $G$ are monotone functors;}
 \item {for all $c\in C$, $d\in D$, $S_{1}(c) \leq S_{1}(GFc)$ and $S_{2}(FGd) \leq S_{2}(d)$;}
 \item {for all $c\in C$, $d\in D$, $S_{2}(FGFc)=S_{2}(Fc)$, and $S_{1}(GFGd)=S_{2}(Gd)$.}
\end{enumerate} 
\end{Corollary}
The proof is repetition of the proof of Theorem \ref{Th.GaloisConnectionEquivalentConditions}.

From Theorem 145 of \cite{CategoryGentleIntroduction} we immediately have transitivity of the Galois connection, namely,
\begin{Theorem}
\label{Theorem_TransitivityOfGaloisConnection}
 The Landauer's connection is transitive, namely, if $F: \mathcal{G}_{1} \rightarrow \mathcal{G}_{2}$, $G:\mathcal{G}_{2} \rightarrow \mathcal{G}_{1}$, $H:\mathcal{G}_{2} \rightarrow \mathcal{G}_{3}$ and $K: \mathcal{G}_{3} \rightarrow \mathcal{G}_{2}$ then if $F\dashv G$ and $H \dashv K$, then also $HF \dashv GK$.
\end{Theorem}
We have also some kind of uniqueness (see Theorem 144 of \cite{CategoryGentleIntroduction}), that is,
\begin{Theorem}
Landauer's connection is 'unique' in the following sense: if $F \dashv G$ and $F \dashv G'$ then $G=G'$. Similarly for the other direction. 
\end{Theorem}

From two Landauer's connected entropy systems we can isolate those parts that are order isomorphic (see Section \ref{Appendix_GaloisConnetion} for definition). We can define an order isomorphism in the following way (see Definition 118 and Theorem 150 of \cite{CategoryGentleIntroduction}): 
\begin{Construction}
Take images of the functors $\Gamma_{1}^{\dashv}=G[\Gamma_{2}]$ and  $\Gamma_{2}^{\dashv}=F[\Gamma_{1}]$ and define new subcategories of posets $\mathcal{G}_{1}^{\dashv}=(\Gamma_{1}^{\dashv},S_{1})$ and $\mathcal{G}_{2}^{\dashv}=(\Gamma_{2}^{\dashv},S_{2})$. Then these last categories are order isomorphic by $F$ and $G$. 
\end{Construction}
The order-isomorphism allows us to introduce classes of equivalences between entropy systems or their subcategories, and therefore introduce in the category of entropy systems (sets of all entropy systems without any arrows apart of identity arrow) the quotient category, whose objects are equivalence classes of order-isomorphism. In addition, order-isomorphism allows us to construct a subsystem of Landauer's connected systems that can be used to implement the other entropy system faithfully.

We can also close\footnote{The closure of $\mathcal{G}_{1}$ (Definition 119 of \cite{CategoryGentleIntroduction}) is an endofunctor $K:\mathcal{G}_{1}\rightarrow \mathcal{G}_{1}$, such that for all $c,c' \in\mathcal{G}_{1}$ we have 1) $c \preccurlyeq Kc$, 2) $K$ is monotone, 3) $K$ is idempotent, i.e., $KKc=Kc$;} the poset $\mathcal{G}_{1}$ (see Theorem 151 of \cite{CategoryGentleIntroduction}) using the functor $K=GF$ for $F \dashv G$, where $F: \mathcal{G}_{1} \rightarrow G_{2}$. This closure gives the biggest subcategory of $\mathcal{G}_{1}$ that can be used to simulate/realize $F[\mathcal{G}_{1}]$.

The above construction of Landauer connection for entropy systems shows that it is 'weakest' relation between them in the sense that it allows only to preserve the entropy ordering between them, since the Galois connection is the 'weakest' connection between posets.

Final part of this section is devoted to explain how the Tab. \ref{Tab1_FormPaperLandauer} is related to the Landauer's connection, and to provide practical tools to indicate this connection.

\begin{Corollary}
\label{Corollary_increaseOrdering}
 Conditions equivalent to the first two conditions of Corollary \ref{Corollary_LandauerFunctorsEquivalentDef} for $F\dashv G$ are
 \begin{enumerate}
  \item {if states $p,p'$ are ordered as follows: $p \preccurlyeq p'$, then $p \preccurlyeq p' \preccurlyeq GF(p')$ and $p \preccurlyeq GF(p) \preccurlyeq GF(p')$; }
  \item {if states $p,p'$ are ordered as follows: $q \sqsubseteq q'$, then $FG(q) \sqsubseteq q \sqsubseteq q'$ and $FG(q) \sqsubseteq FG(q') \sqsubseteq q'$.}
 \end{enumerate}
\end{Corollary}
The Corollary says that the Galois' connection is in general not symmetric operation, and $GF$ functor maps a state to the 'higher or equal' state. Likewise, $FG$ maps a state to the 'lower of equal' state.

Order-embedding and order-isomorphism conditions are given by the following standard results from the Galois connection theory \cite{CategoryGentleIntroduction,CategoryTheoryForWorkingMathematicans}
\begin{Lemma}
\label{Lemma_Surjectivity}
 The following conditions are equivalent for $F\dashv G$
 \begin{enumerate}
  \item {$GF(p) =p, \quad \forall p \in \Gamma_{1}$,}
  \item {$F$ is surjective,}
  \item {$G$ is injective.}
 \end{enumerate}
\end{Lemma}
In general $T=GF$ is a closure functor, however
\begin{Corollary}
\label{Corollary_OrderIso}
 If $FG = Id_{\Gamma_{2}}$ and $GF=Id_{\Gamma_{1}}$ then $\Gamma_{1}$ and $\Gamma_{2}$ are order-isomorphic.
\end{Corollary}

Finally, we need the definition of reversibility, which mimics thermodynamic reversibility of a process, namely 
\begin{Definition}
 An entropy system map, that is a poset map $f:\Gamma \rightarrow \Gamma$ is reversible at $p \in \Gamma$, if  $p=f(p)$, that is $S(p)=S(f(p))$, i.e. $f$ at $p$ preserves entropy. Otherwise $f$ is irreversible at $p$. 
\end{Definition}
Note that reversibility is connected with map and the state on which it acts. Such maps in thermodynamics are called reversible processes and in logic/computing reversible operations.
 
We are ready to formulate the main theorem that connects classical Landauer's principle given by Tab. \ref{Tab1_FormPaperLandauer} with the Landauer's connection
\begin{Theorem}
\label{Theorem_Main}
 For two entropy systems $\mathcal{G}_{1}=(\Gamma_{1}, \preccurlyeq)$ and $\mathcal{G}_{2}=(\Gamma_{2}, \sqsubseteq)$, and functors $F:\mathcal{G}_{1}\rightarrow \mathcal{G}_{2}$ and $G:\mathcal{G}_{2}\rightarrow \mathcal{G}_{1}$, we have following possibilities for Landauer-Galois' connections
 \begin{enumerate}
  \item {
 \begin{tabular}[!h]{|c|c|c|}
 \hline
 Possibilities &  \begin{tabular}{@{}c@{}}$\Gamma_{2}$ \\ reversible\end{tabular}  & \begin{tabular}{@{}c@{}}$\Gamma_{2}$ \\ irreversible\end{tabular} \\ \hline
 $\Gamma_{1}$ reversible & YES  & YES \\ \hline
 $\Gamma_{1}$ irreversible & NO & YES \\ \hline
\end{tabular}
    for which  $F\dashv G$, \\
  }
  \item {
 \begin{tabular}[!h]{|c|c|c|}
 \hline
 Possibilities &  \begin{tabular}{@{}c@{}}$\Gamma_{2}$ \\ reversible\end{tabular}  & \begin{tabular}{@{}c@{}}$\Gamma_{2}$ \\ irreversible\end{tabular} \\ \hline
 $\Gamma_{1}$ reversible & YES  & NO \\ \hline
 $\Gamma_{1}$ irreversible & YES & YES \\ \hline
\end{tabular}
    for which  $G\dashv F$, \\
  }
 \item {
 \begin{tabular}[!h]{|c|c|c|}
 \hline
 Possibilities &  \begin{tabular}{@{}c@{}}$\Gamma_{2}$ \\ reversible\end{tabular}  & \begin{tabular}{@{}c@{}}$\Gamma_{2}$ \\ irreversible\end{tabular} \\ \hline
 $\Gamma_{1}$ reversible & YES  & NO \\ \hline
 $\Gamma_{1}$ irreversible & NO & YES \\ \hline
\end{tabular}
    for which $F, G$ are order-embeddings; If the functors are surjective, then they are order-isomorphisms.
  }
 \end{enumerate}
\end{Theorem}

\begin{Proof}
 For the first claim let us take a state $p \in \Gamma_{1}$ and a reversible map $f:\Gamma_{1}\rightarrow \Gamma_{1}$ that gives $p' =f(p)$, such that $p=p'$ (that is $S_{1}(p)=S_{1}(p')$, where $S_{1}$ is the entropy in $\mathcal{G}_{1}$.) Using the first claim of Corollary \ref{Corollary_increaseOrdering} we have $p=f(p)\preccurlyeq GF(f(p))$ and $p \preccurlyeq FG(p) \preccurlyeq GF(f(p'))$ and $F,G$ are monotone functors. Therefore $F$ can map $p=p'$ into $F(p)=F(p')$ or $F(p) \sqsubseteq F(p')$, $F(p) \neq F(p')$, i.e. $Ff$ is reversible or irreversible process.
 Irreversible process in $\mathcal{G}_{1}$ is always mapped (functors are monotone) into irreversible process - third row of the table.

 Similar argument in opposite direction holds for the second case.
 
 For the third case the functors $F$ and $G$ must map reversible maps to reversible maps and irreversible maps to irreversible ones. Therefore they preserve ordering so they are order-embeddings onto images, see Lemma \ref{Lemma_Surjectivity}. If in addition $F$ and $G$ are surjective functors, then from Corollary \ref{Corollary_OrderIso} it results that they are order-isomorphisms.
\end{Proof}
The above theorem is a simple tool that helps to detect the presence of the Landauer's connection and their direction.

In the next section examples of the Landauer's connection, including original Landauer's principle will be given. 

\section{Examples}
This section presents a few examples of various level of details of interplay between entropy systems and Galois connection that we call the Landauer's connection. 
We start from simple yet imaginative toy example.

\subsection{Toy example}
This example will be motivated by a simple example (Example 1.80) of the Galois connection from \cite{SpivakSketches}. 

Consider two entropy systems $\Gamma_{1}=(\mathbb{R}_{\geqslant 0}, S)$ and $\Gamma_{2} = (\mathbb{N}_{\geqslant 0}, S)$, where the entropy in both cases is given\footnote{Different function can be selected, e.g., the floor function $S(x)=\lfloor x \rfloor$. In this case different ordering on real numbers is used, however derivations are similar.} by the identity $S(x)=x$. This choice of entropy agrees with standard ordering on natural and real numbers.

One can see that the system $\Gamma_{1}$ has higher cardinality of states than $\Gamma_2$. We will consider the Galois extension in both ways as in \cite{SpivakSketches}. In the considerations these posets are treated as categories on their own, and therefore functors are simple monotone functions.

\textbf{Case 1.} 
Consider $F: \Gamma_{1} \rightarrow \Gamma_{2}$ defined as $F(z) = \lceil \frac{z}{3} \rceil$ and $G: \Gamma_{2} \rightarrow \Gamma_{1}$ given by $G(z) = 3z$.  We have $F \dashv G$ since it fulfils (\ref{Eq.GaloisConnectionG}) i.e.
\begin{equation}
 \left\lceil \frac{x}{3} \right\rceil \leq y \quad \Leftrightarrow \quad x \leq 3y.
\end{equation}
Let us consider a few processes on $\Gamma_{1}$ and related processes in $\Gamma_{2}$ induced by the Galois connection:
\begin{itemize}
 \item {Consider now the following map $f: \Gamma_{1} \rightarrow \Gamma_1$ given by a simple shift $f(z) = z+0.2$. Take $x=1 \in \Gamma_{1}$ for which $S(x)= 1$. Then $\bar{x}=f(x) = 1.2$ and $S(f(x)) = 1.2$ and therefore process $x \rightarrow \bar{x}$ is irreversible (entropy increases). We have $y=F(x) = 1$ with $S(y)=1$, and $\bar{y}=F(\bar{x})=Ff(\bar{x}) = 1$ with $S(\bar{y})=1$, and therefore, the irreversible process in $\Gamma_{1}$ is mapped by $F$ to reversible process  on the level of $\Gamma_{2}$.}
 \item {Take the same map $f(x) = x+0.2$ with initial point $x=2.9$. It gives $\bar{x} = f(x) = 3.1$ and therefore $S(x)=2.9$ and $S(\bar{x})= 3.1$ - irreversible process in $\Gamma_{1}$. Using functor $F$ we get $y=F(x) = 1$ and $\bar{y}=F(\bar{y}) = 2$. Therefore irreversible process in $\Gamma_{1}$ is mapped to irreversible process in $\Gamma_{2}$. }
 \item {If we take $f(x)=x$ then reversible (trivial) process in $\Gamma_{1}$ is mapped to reversible process in $\Gamma_{2}$}
 \item {No irreversible process in $\Gamma_{2}$ can be realized by a reversible process in $\Gamma_{1}$.}
\end{itemize}
Summing up, proposed Galois connection gives case 1 from Theorem \ref{Theorem_Main}.

\textbf{Case 2.}
We now take $F: \Gamma_{2} \rightarrow \Gamma_{1}$ defined as $F(x) = 3x$ and $G: \Gamma_{1} \rightarrow \Gamma_{2}$ given by $G(x) =  \lfloor \frac{x}{3} \rfloor $.
This also defines the Galois connection $F \dashv G$ as it is easily checked. We have the following examples of processes:
\begin{itemize}
 \item {The process in $\Gamma_{1}$, e.g. the shift $f(z)=z+3$ that irreversibly maps $x=6$ to $\bar{x}=9$ on the level of $\Gamma_{2}$ gives the map from $y=G(x)=2$ into $\bar{y}=G(\bar{x})= 3$ which is also irreversible.}
 \item {For the irreversible shift $f(z)=z+0.1$ on $\Gamma_{1}$ that maps $x=2$ to $\bar{x}=2.1$ we have reversible (identity) process in $\Gamma_{2}$ that maps $y=G(x)=0$ to $\bar{x}=G(\bar{y})=0$ that is obviously reversible.}
 \item {Identity (reversible) process in $\Gamma_{1}$ is trivially mapped into reversible process in $\Gamma_{2}$.}
 \item {There is no mapping of an irreversible process on $\Gamma_{1}$ to a reversible process in $\Gamma_{2}$.}
\end{itemize}
In summary, the case 2 of Theorem \ref{Theorem_Main} is restored.

\textbf{Case 3.}
For an example of the case 3 of Theorem \ref{Theorem_Main} consider identity mapping $F=Id_{\Gamma_{1}} = G$ between two copies of $\Gamma_{1}$.

First two examples show how a system with 'bigger multiplicity of states' is Galois connected with a system with 'smaller number of states' and fulfils the claims of Theorem \ref{Theorem_Main} relating reversible or irreversible processes between these two  systems. Similar principle can be used in description of memory chip where two logical states can be realized by some complicated sets of physical states and their internal transitions that realize binary operations. This idea will be used in the next subsection.

\subsection{Landauer's functors and Maxwell's demon}
In this subsection we describe how the above abstract language can be applied to description of original Landauer's principle of irreversible computations. Then, well-known the Maxwell's Demon paradox will be presented using Landauer's connection. This is use of a new and more powerful language to the known solution described in details in \cite{LandauerExplained}.

We would like to stress again that this is not a new solution for the problem, which was solved already. Instead, it is reformulation of the problem in the new abstract language of the Landauer's connection which in our opinion gives clearer and more uniform description of the problem. The thermodynamic details are hidden in the details of the Galois connection and their manifest themselves in heat emission during irreversible computation.

\begin{figure}
\centering
 \includegraphics[width=0.8\textwidth]{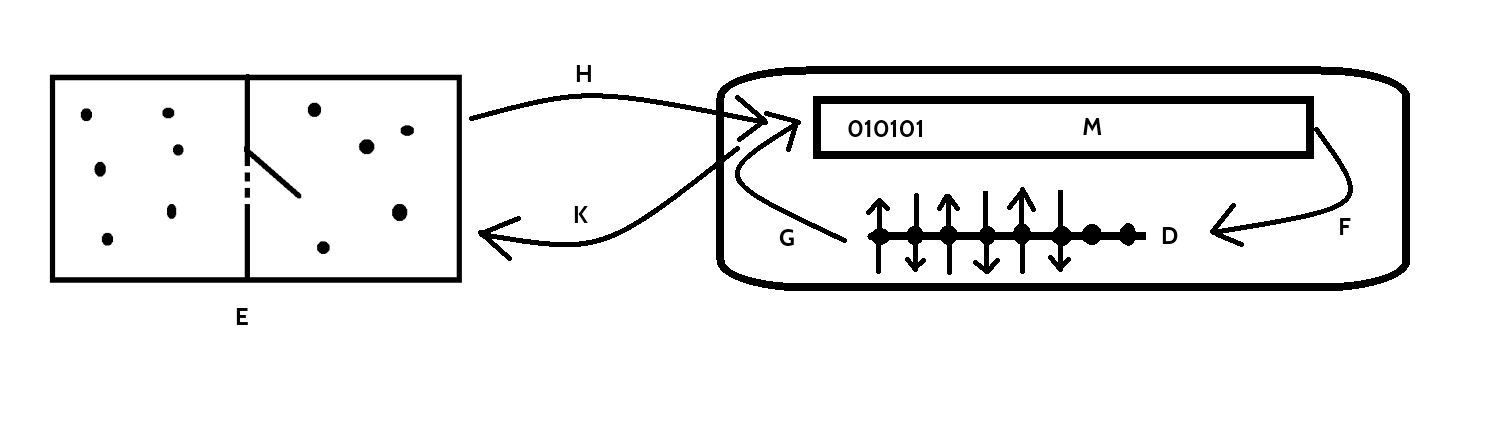}
 \caption{The Landauer's connection between box with ideal gas $E$, memory $M$ of the Maxwell's Demon and its physical realization $D$ in the Maxwell's Demon experiment.}
 \label{Fig.MaxwellsDemon}
\end{figure}

Let us first explain the classical Landauer's principle in therms of the Landauer's connection introduced in the previous subsection. Let us consider first computer memory $M$ that bases on binary logic, and its implementation using some physical system $D$. In both cases they are entropy systems (see Fig. \ref{Fig.MaxwellsDemon}). 
We can therefore build posets using entropy as ordering, namely construct $(M, S)$ and $(D,\bar{S})$, where $S$ and $\bar{S}$ are corresponding entropies.

Since the relation between logical part and physical realization of the memory is described by Tab. \ref{Tab1_FormPaperLandauer}, therefore from Theorem \ref{Theorem_Main}, there is a Ladnauer's connection $F \dashv G$ between functors $F:M \rightarrow D$ and $G: D \rightarrow M$. The details of the functors depend on the implementation, however they are maps between logical states and corresponding physical states in the memory implementation. If there is irreversible operation on the memory $M$ given by a function $f:M \rightarrow M$, then it induces irreversible operation on the device $D$ given by $Ff: FM \rightarrow FM$, and this, by the Second Law of Thermodynamics generates heat that is expelled to the environment. The amount of emitted heat depends on the realization (i.e. on properties of $F$ and $G$), however Landauer showed the lower bound for it, namely, $k_{B}T\ln 2$.

\begin{figure}
\centering
 \includegraphics[width=0.8\textwidth]{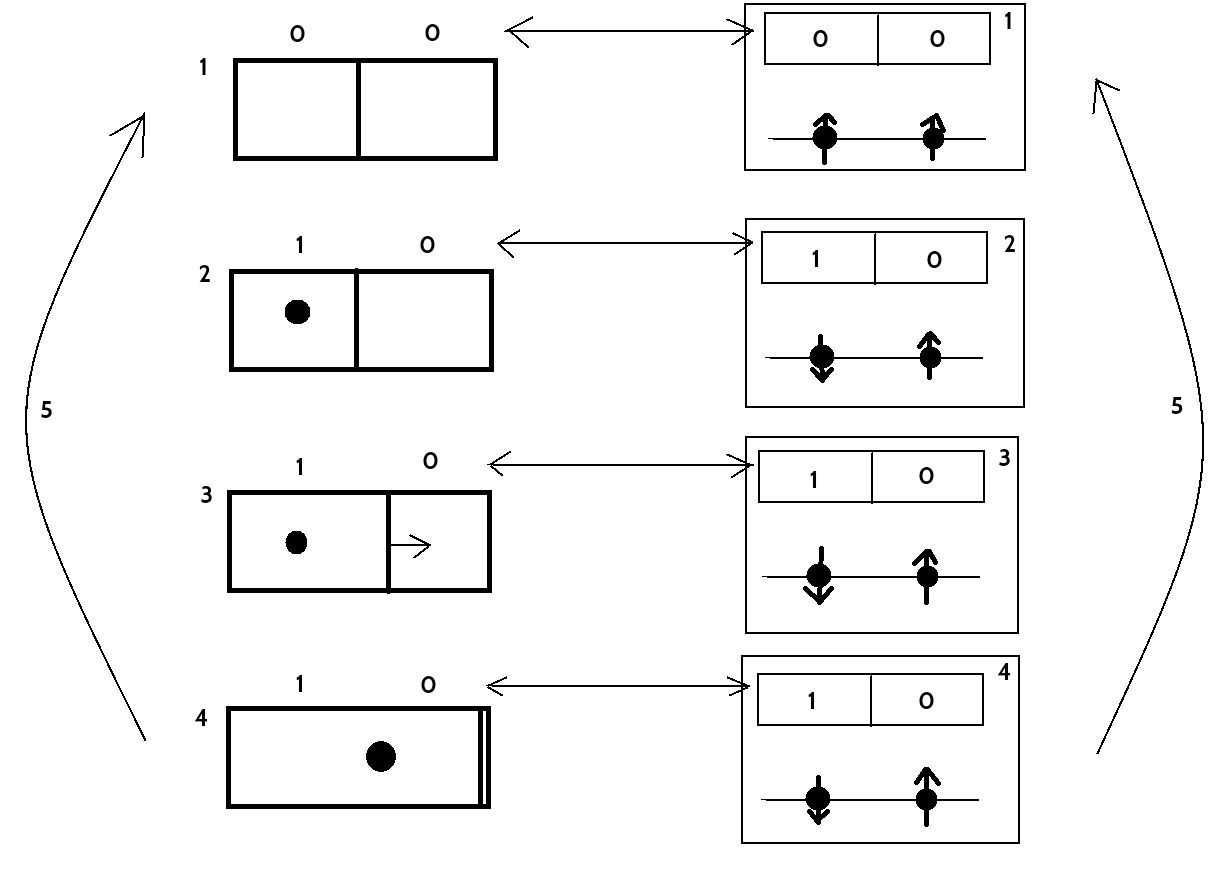}
 \caption{Maxwell's Demon experiment with a single particle and movable partition. On the left there is a box with movable partition and on the right corresponding memory state.}
 \label{Fig.MaxwellsDemondetails}
\end{figure}

In the next part of our considerations this memory $M$ is used as a Demon's memory in the Maxwell's Demon 'paradox'. In the experiment there is the thermodynamic system - the box with an ideal gas, and the partition that can selectively be opened - part $E$. It is connected with the memory $M$, which saves informations on separation of gas particles depending of their kinetic energy, e.g. high kinetic energy particles are collected in the left and low energy particles in the right chamber. The functors $K:E\rightarrow M$ and $H: M \rightarrow E$ are relations between information on localization of the particle in $E$ and its logical description in $M$ (not in $D$). For details consider Fig. \ref{Fig.MaxwellsDemondetails} where simplified Maxwell's Demon experiment (due to Le\'{o} Szil\'{a}rd \cite{Szilard}) with one gas particle is presented.
We will describe every step in the cycle. Two-bit description of the logical state of the system was selected for better understanding - $1$ means that the particle is localized in the left ($10$) or the right ($01$) part of the box.
\begin{enumerate}
 \item {In this state there is no information on localization of the particle, and therefore information entropy $S_{i}=2$ as the state is the mixture of two states $01$ and $10$. This state is associated to $00$ bit description (reset) and transferred to the memory.}
 \item {Particle was localized (for example) in the left chamber ($10$) so the information entropy is now $S_{i} = 1$. This state is transferred into memory, where  reversible operation (e.g. $NOT \otimes Id$) is performed that change $00$ into $10$.}
 \item {Partition starts to move freely. There is adiabatic decompression of a single particle gas. State of the memory is the same as in the previous step.}
 \item {Partition is pushed maximally to the right. The work done by the particle is $W = k_{B}T\int_{V/2}^{V}\frac{dV}{V}=k_{B}T\ln 2$, where $V$ is the volume of the box, $k_{B}$ is the Boltzmann constant, and $T$ is the temperature. Since no heat flow was present, thermodynamic entropy is still constant and the internal energy of the gas decreased. New cycle will start.}
 \item {This transition is the restart of the cycle. The partition is placed in the middle of the box, and therefore information on localization of the particle is lost. Information entropy is now $S_{i}=2$. State become $00$ and it is correlated with the state of the memory - irreversible operation (e.g. $f(x) = 00 \quad AND \quad x$) is performed on the memory, which results in expelling, via Landauer's principle, at least $k_{B}T\ln 2$ of heat form its physical part $D$. Cycle repeats.}
\end{enumerate}
One can note that $K$ and $H$ are in fact isomorphisms and they connect the state of the knowledge on the particle position and state of the memory. 

Irreversible operation on the memory $f: M \rightarrow M$ is transferred via the Landauer's functor $F$ into irreversible operation $Ff$ on its physical implementation. That results in expelling heat and preserves the Second Law of Thermodynamics for the whole system in every cycle. 

If a bigger memory would be used for storing information on the particle localization in a few cycles, then its erasing would  expel at least multiple of $k_{B}T\ln 2$ of heat from $D$ part and preserved the Second Law of Thermodynamics after these cycles. In this case the end of the cycle is marked by erasing of the memory and not the thermodynamic cycle in $E$ part of the system.

It is interesting to note that the irreversible process $f$ can be transferred to $E$ as $Kf$, however it produces heat only in $D$ part of the system (however corresponding changes of entropies appear in $E$, $M$ and $D$). This explains how the Second Law of Thermodynamics and entropy changes can be transferred from $D$ to $E$, i.e., irreversible operation $f$ corresponds to irreversible operation $KGFf$ on $E$.

\subsection{DNA computation}
In this and the next subsection a sketch of application of the Galois connection to biochemical and biological systems will be presented. Due to large scale of complexity of such systems the description is not detailed and many statements can be treated rather as research hypotheses than firmly stated claims. We believe however that promoting the Galois-Landauer principle to a general principle justifies its formulations in abstract language of category theory, which makes it possible to trace it also in biochemical systems and living organisms.

Every living cell contains 'a computer' that operates on complicated chemical principles using DNA (deoxyribonucleic acid), and controls every aspect of the cell. In recent years such principles were used to implement efficiently some computationally difficult algorithms thanks to enormous parallelization achievable by this approach. For an excellent overview of this subject see \cite{DNAComputingOverview} and references therein.

In short, DNA as a storage of genetic informations consist four bases: A (adenine), G (guanine), C (cytosine) and T (thymine). Single DNA strand can be considered as a list composed with these four letters. The second strand can be connected using the following complementary connection rules: A-T and G-C. Therefore single strand consists the same amount of information as a double one. In addition, the direction (polarity) of the strand is marked by chemical compounds named $3'$ and $5'$. An example of a short fragment of a DNA strand (called oligonucleotide) is $3'$ACTGTA$5'$. 

Operations on such structure are controlled by changing physical properties of environment (temperature that, e.g., decides if the DNA double helix decouples into individual strands -melting, or combines single strands into double list - annealing) or chemical properties (especially by adding specially designed enzymes that perform various operations on short pieces of DNA \cite{DNAComputingOverview}). In computation some external, and not present in living organisms, methods like gel electrophoresis are used \cite{DNAComputingOverview}.

Following \cite{DNAAlgebraic}, the most effective representation of bases that is unchanged after changing polarity and the respects complementarity is the three-bit association: A-000, C- 010, G-101, T-111. One can then represents basic operations in algebraic form \cite{DNAAlgebraic}.

DNA computing and DNA processing in a living cell is a complicated sequence of chemical reactions in some environment usually without sharp borders, and therefore we present only a rough idea how it can be connected with the Galois connection. 

\begin{figure}
\centering
 \includegraphics[width=0.5\textwidth]{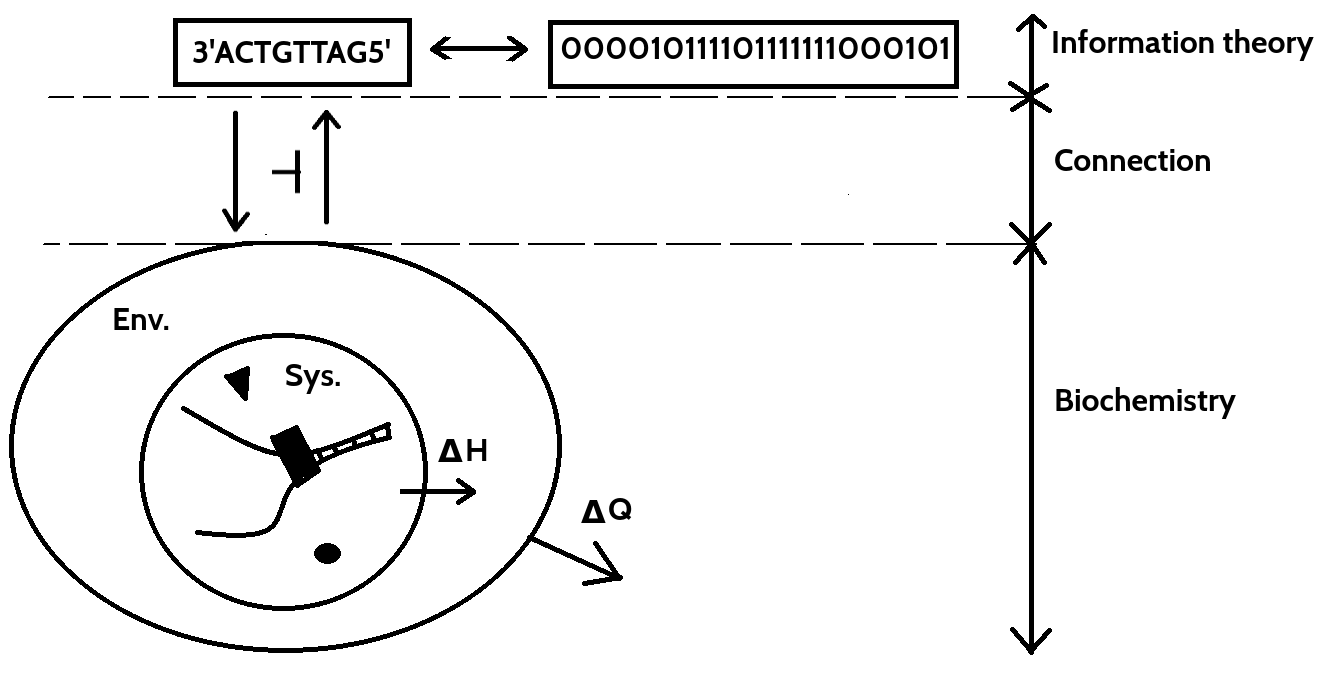}
 \caption{Schematic structure of computation in DNA. On the level of computation realm (Information theory) there is information encoded in DNA strand. It is Galois connected with biochemical system which realizes computations by means of chemical reactions. This system contains an Environment (Env.) and embeds inside the System (Sys.) with DNA, chemical elements and enzymes, where actual computation takes place. The Environment interacts with the System for conducting specific chemical reactions that realizes logical operations. The System and Environment overall fulfils The Second Law of Thermodynamics and therefore the total entropy can remain constant or increase, i.e. $\Delta S_{Env} + \Delta S_{Sys.} = \Delta S_{T} \geq 0$. Defining the enthalpy change (dispersed heat of the System) as $\Delta H = -T\Delta S_{Sys.}$ and the Gibbs free energy change as $\Delta G = -T\Delta S_{T}$ on gets the famous equation $\Delta G = \Delta H - T\Delta S_{Sys}$. All reactions in the system are spontaneous if $\Delta S_{T} >0$, that is $\Delta G <0$. This is the principle of interaction between the System and the Environment.}
 \label{Fig.DNAComputing}
\end{figure}

On the logical level (see Fig. \ref{Fig.DNAComputing}), as it was described above and in \cite{DNAAlgebraic, DNAComputingOverview}, there is well-formalized set of operations on logical representation of DNA state. In this system various flavours/notions of entropy that capture different levels of computations can be described - from the Shannon entropy \cite{DNAInformationEntropy}, through block entropy \cite{DNABlockEntropy} and topological entropy \cite{DNATopologicalEntropy} among others.

On chemical (realization) level (see Fig. \ref{Fig.DNAComputing}) due to enormous complexity the reactions \cite{VitalQuestion} in the system with DNA cannot be decoupled from the enclosing environment in which the system is embedded (e.g. living cell, biochemical reactor). The boundary of the environment depends on how complex computation is, e.g., for 'small' computation it can be nucleus of the cell or its membrane (or the test-tube in which compounds are for in vitro computations). However for long and complex computations probably the environment in which the cell is living including other cells would be a good choice. Some hints on selecting boundary of the environment result from thermodynamics - boundary should be such a (natural or artificial) barrier/closed surface that the total entropy inside it should always increase or remains constant for the whole process. In other words it should be minimal boundary of volume in which the Second Law of Thermodynamics is fulfilled for all time of the process.

This whole composed system and environment have to obey the Second Law of Thermodynamic. Interaction between the System and the Environment drive the System with DNA to perform some specific reactions by changing physical and chemical properties of the System. That leads to change in the Gibbs free energy (Fig. \ref{Fig.DNAComputing}) that determines direction of the reactions in the DNA system. 

Similarity with the model of physical memory described above suggests that such relation between information stored in DNA and various logical operations from logical side, and their chemical realization by the system and environment qualify to describe them in terms of the Galois-Landauer's connection. If this hypothesis is true then it shed some light on basic principles of life on elementary level. This is reasonable hypothesis since such complicated chemical reactions (whatever optimized by evolution process) should always increase total entropy (nontrivial reversible and irreversible computations should be realized by irreversible chemical processes that increase total entropy). In addition, every irreversible computation on logical level, via hypothetical connection, release the Landauer's heat to the environment.

The model of DNA as a memory modelled using the Galois connection is also vital in view of recent experiment that showed how to encode large amount of data (short movie) in DNA of living organism \cite{DNAMovie}.   

In this example we motivated that the Galois connection is a model for operation on single genetic bases and their chemical realization. In the next example we will present that the similar structure should exists on the level of genes (conglomerates of genetic information that encode proteins structures) and animal species in the Tree of life.

\subsection{Is 42 the meaning of life?}
\label{subsection_42}
The provocative title of this subsection refers to the fiction book \emph{The Hitchhiker's Guide to the Galaxy} by Douglas Adams. He describes an advanced civilization that designed planet-size computer (Earth) with a 'biological component' that should answer the ''Ultimate Question of Life, The Universe, and Everything''. This fictional idea surprisingly well resembles the following construction. 

The hypothesis on existence of the Galois in the previous example, if true, shows that that on the basic level, life is a computation process on chemical components. On larger scales the Galois connecting  can also conjugate expression of gene pool and animal species interconnections. This vague idea was described in \cite{SpivakSketches}, Example 1.84, and it is worth of citing here for completing the picture. Due to complexity of biological examples only rough idea will be presented. Some general idea on the level of such complexity is presented in \cite{VitalQuestion} and reverences therein.

Following \cite{SpivakSketches}, let $(P, \subseteq )$ is the poset describing possible animal populations with inclusion $p \subseteq q$ if the animal species $p$ is also the animal species $q$ in the sense of specificity on the Tree of life (see Example 1.51 in \cite{SpivakSketches}). Moreover, the inclusion of species induce some kind of entropy that can be used to measure various changes in the Tree of Life. 

The other poset $(G , \leq )$ describes gene polls and the ordering has the following meaning: $ a \leq b$ when the gene pool $b$ can be generated by the gene pool $a$. The ordering also can be used to define some kind of entropy that measures ordering or information loss on the level of gene pools.

The Galois connection is defined by two monotone functions (functors, when posets are treated as categories on their own) \cite{SpivakSketches}. The first one $i: P \rightarrow G$ sends each population to the gene pool that defines it. The second functor $cl: G\rightarrow P$ sends each gene pool to the set of animals that can be obtained by recombination of the given gene pool. Then $i \dashv cl$ in this very broad sense.

This example involves structure of the whole population of organism on Earth and their genetic information as 'a database' for processes that describe computations (evolution). Since connected systems are not of thermodynamic origin therefore no heat during the irreversible process is expelled (it is even difficult to define such quantity not having a definition of temperature in the model). Still, operating on not too strict level, the process of evolution can be decomposed, using the hypothesis from the previous chapter, into chemical reactions. Since evolution is a long term process therefore the environment for computation would involve the whole Earth and all organisms in which biochemical reactions take place. This is unrealistic model, and therefore, usually focusing on small piece of systems and their interaction with environment is more reasonable approach \cite{BaezEntropy1, BaezEntropy2}. However, this abstract level motivates the introduction of 'the heat of evolution' - Landauer's heat of all chemically realized bio-computations that were (and still are) realized in the process of evolution.

This idea can serve as a model of life on Earth, however the way to state it precisely requires mathematical biology, biochemy and taxonomy to be developed on very detailed and precise level not available currently. Only then exact definition of Galois functors can be provided. Moreover, this slightly modified example of the Galois connection from \cite{SpivakSketches} gives hints how to develop entropy measures consistent (preserved by the connection) between biochemical, genetic and taxonomy levels provided that the details of the Galois connections are known.

The situation in modelling such structure is somehow simpler in evolutionary robotics \cite{EvolutionaryRobotics} where virtual environment, that resembles some features of physical environment in which robot operates, is used to simulate artificial evolution that is optimization of robot shapes. Genetic code in this case is a set of optimized parameters of robot (e.g. the length of the legs of a robot or its neural network design) and environment is represented by some multidimensional function called fitness function \cite{EvolutionaryRobotics-FitnessFunction}. Virtual evolution is in fact multidimensional optimization process that is aimed to find minimum of fitness function (which can be globally shifted to have value equals to $42$). This minimum (which can be non-unique) describes optimal fitness in a given environment for the robot construction with respect to optimized parameters. Due to larger 'rigidity' of environment (usually fitness function does not change or vary slowly) than in biological situation (when environment changes abruptly and contains other interacting specimens), the construction of the Galois connection between robots genetic poll and their construction features, and related entropies (in analogy to the presented above biological example), should be easier. Possibility and details of such construction deserves another paper.

Summing up abstract discussion from the last two subsections, on every level of life there is a pattern which resembles the Galois connection. This pattern can be a hint for an emergent phenomena in biology.

\section{Discussion}
The Galois connection was originally invented to model the relationship between semantic and syntax \cite{GaloisAndSemanticSynctatic} of mathematical theories - the relation between set of axioms and classes of models that implements such axioms. It is a surprising coincidence that the same structure exists in systems governed by entropy and describing realization/simulation of one system by the other. It can be understood in the terms that one simulated system gives a set of axioms that have to be met and another simulation system is a model that is used to map the behaviour of the first simulated system. 

There is also an even more important interpretation in the view of entropy as the loss of information   when we formulate a physical model from uncertain measurement data \cite{EntropyLychagin}. In this context this information loss can be propagated via Landauer's connection to the other related systems. 

Likewise, from the Maxwell's Demon example we observe that the Second Law of Thermodynamics propagates along the Landauer-Galois' connection and for a proper thermodynamic description of the system, all connected parts, must considered together.

Landauer's connection and information interpretation of entropy, bear some relationships to MUH (Mathematical Universe Hypothesis) articulated by Max Tegmark \cite{TegmarkMathematicalUniverse}. The observations in this paper may be reminiscence of computational/information principles on which our universe is based. The Landauer's connection can be relation between encoding physical laws and its 'equivalence class of descriptions' \cite{TegmarkMathematicalUniverse}.

One of the basic questions that arise from the Landauer's connection is whether there are (ultimate) computations that are not connected with physical realizations, or whether more generally, is every entropy system connected with some other entropy system, i.e. we can group entropy systems in pairs without no one left out?

From the above connection some insight on the black hole entropy and information loss after crossing the event horizon could be anticipated when the realization of such information via the Landauer's connection is taken into account. This deserves another paper.

\section{Conclusions}
In the paper it was shown that the well-known and experimentally confirmed Landauer's principle is a reflection of a general connection between entropy systems, called the Galois connection. This connection was cast into systems where entropy exists naturally. The result of entropy-induced ordering was used to provide poset structure in such systems (defined as entropy systems) and this gives link between Galois connection and Landauer's principle. 

The original Landauer's principle was restricted to the connected information-physical systems, however, the general form of the Landauer's connection presented in this paper can be applied to every pair of systems with some kind of mapping (realization/implementation) that connects states of both systems. Then Landauer's connection preserves entropy structure, and therefore preserves poset structure on the state-spaces of the systems that is induced by entropy.

In the discussion it was argued that this connection can be a reminiscent of more fundamental principle which our universe is based on.

Ladauer's connection presented in the paper can be used as a backbone of category theory model that describes relation between two entropy systems.

\section*{Acknowledgments}

I would like to thanks those who made me interested in category theory and abstract approach to science - those who I met in person or know only by their magnificent books and papers. This research was supported by the GACR grant 17-19437S, and the grant MUNI/A/1138/2017 of Masaryk University.

\appendix




\end{document}